\documentclass[aps,12pt,final,notitlepage,oneside,onecolumn,nobibnotes,nofootinbib,superscriptaddress,noshowpacs,centertags]{revtex4}
\usepackage{amsmath,amssymb,epsfig,placeins,subfigure,wrapfig,indentfirst,makeidx}
\usepackage[T2A]{fontenc}
\usepackage[utf8x]{inputenc}
\usepackage{ucs}
\usepackage{amsmath} 
\usepackage{mathtext}
\usepackage{amsfonts}
\usepackage{upgreek}
\usepackage{amsfonts,amssymb}
\usepackage{color}
\usepackage{graphicx}
\usepackage{mathrsfs}
\usepackage{dsfont}

\begin{document}

\date{\today}

\title{Occurrence of a Turning Point in the Dynamic Solution for a Reissner–Nordström Black Hole}

\author{Alexander Shatskiy}
\affiliation{shatskiyalex@gmail.com}

\maketitle

{\bf Abstract:} 

A self-consistent exact solution for a Reissner–Nordström black-and-white hole formed as a result 
of accretion has been considered. 
Prior to the formation of a black-and-white hole, there is a bulk charged 
sphere at the center of the system. The occurrence of a turning (bounce) point in the newly formed black-and-white hole is investigated. 
The occurrence of a turning point was investigated by an example of solution 
for the accretion of a neutral spherical dust shell. 
The model equations are written with allowance for the cosmological $\Lambda$-term. 
Within the model under consideration, both the black-and-white hole and its turning 
point are formed in the already existing Universe; therefore, the black-and-white hole in this model is not <<eternal>>.

\section{NTRODUCTION}
\label{s_int}

The existence of black holes (BHs), as well as the mechanisms of their formation, have long been considered as investigated. 
The mathematical solutions for real (rotating) Kerr BHs and charged Reissner–Nordström BHs show the existence of a turning point.
The turning (or bounce) point is considered to be a point at which the free-fall trajectory of a particle changes from the direction along which the particle
radial coordinate decreases to a direction along which this coordinate increases. 
Here, of interest are specifically the turning points located under both black hole horizons; i.e., under the Cauchy horizon: in the internal R-region of space.

The existence of internal turning point (turning point below) in the steady-state solution for BH is related to the existence of black-and-white hole rather
than a conventional BH. 
However, the mechanism of turning-point occurrence in the self-consistent solution of collapsing matter has not yet been studied 
(see, e.g., \cite{Thorn1977}, \S 34.6). 
Therefore, the models describing the occurrence of black-and-white holes still contain more questions than answers. 
It is agreed that black-and-white holes can be only <<eternal>> BHs, i.e., the ones formed jointly with the creation of the Universe.
Therefore, one of the purposes of this study was to justify the model in which a black-and-white hole itself and its turning point can be formed in the already
existing Universe. 
Note that the white hole is formed in the new expanding universe rather than in the same Universe in which the black hole is formed.

We will consider a self-consistent exact solution for a spherically symmetric Reissner–Nordström black-and-white hole during its formation. 
Prior to the formation of a black-and-white hole, there is a bulk charged sphere at the center of the system. 
The occurrence of a turning point was investigated by an example of exact solution for the accretion of a thin neutral spherical dust shell. 
The model equations will be written with allowance for the cosmological $\Lambda$-term.

Previously many researchers have studied the dynamics of spherical dust shells in a centrally symmetric electric field (see, e.g., \cite{Tretyakova2012, Shatskiy2010, Berezin1}). 
In particular, since the topology of Reissner–Nordström BH was
investigated by plotting Carter–Penrose diagrams in \cite{Berezin1}, we omit these diagrams here and refer the reader 
to \cite{Berezin1} for the information on this issue. 
The topology of real rotating BHs was investigated by the author in \cite{Shatskiy2020}.

\section{DESCRIPTION OF THE MODEL}
\label{s_desc}

Let there be no BH and horizons at the initial instant. 
The matter distribution at the initial instant is described by the following model.
There is a sphere at the center of the system, which has an electrical charge $q$, radius $r_q$ , and mass $m_q$. 
Let the mass $m_q$ be insufficient for the formation of a BH and gravitational collapse of the sphere.

Let also a spherical dust shell with a mass ${m_{dust}}$ be at rest on some radius $R_1$ at the initial instant. 
During further evolution this dust shell begins to fall freely towards the center of the system. 
The mass ${m_{dust}}$ turns out to be sufficient for the further formation of a Reissner–Nordström BH before reaching the shell radius $r_q$. 
Let the total mass of the system be M. The BH horizon radius $r_h^{+}$ and the Cauchy horizon radius ${r_h^{-}}$, corresponding to the charge $q$ and mass ${M}$, 
can be written in the Reissner–Nordström coordinates as\footnote{In this study, if not specified separately, we use the theoretical system of units, 
in which the speed of light and gravitational constant are, respectively, ${c=1}$ and ${G=1}$.}: 
\begin{equation}
r_h^{\pm} = M \pm \sqrt{M^2 - q^2} 
\label{r_pm}\end{equation}
For convenience we introduce the charge-to-mass ratio so as to satisfy the relations
\begin{equation}
r_h^{-}:=\beta^2 r_h^{+} \, ,\quad \beta = const \leq 1 \, .
\label{r_p}\end{equation}
According to (\ref{r_pm}), the necessary conditions for this situation are
\begin{equation}
q = \kappa M \, ,\quad \kappa := \frac{2\beta}{1 + \beta^2} \leq 1 \, .
\label{r_qM}\end{equation}
We introduce also a factor $\gamma$ to link the charge and mass with the radius $r_q$:
\begin{equation}
r_q = \gamma r_h^{-} = \gamma\beta^2 r_h^{+} \, ,\quad \gamma = const < 1 \, .
\label{r_q1}\end{equation}
Obviously, the factor $\gamma$ should to smaller than unity, because both BH horizons must arise before the possible collision of the falling dust shell with the charged sphere; 
i.e., the condition ${r_q<r_h^{-}\leq r_h^{+}}$ must be satisfied.

It follows from the above expressions that
\begin{equation}
r_q = \beta\gamma q = \beta\gamma\kappa M \, , \quad r_h^- = \beta q\, , \quad r_h^+ = q/\beta \, .
\label{r_q2}\end{equation}
Below we will prove the possibility of existence of turning point ${r_{t}}$, which arises in solution in the gap between the radii $r_q$ and ${r_h^{-}}$.

To proceed further, we should solve the problem with limitation on the sphere size $r_q$. 
Here, the main limiting factor is the electric field strength $E$. 
The maximum strength, ${E=q/r^2}$, should be below the vacuum breakdown: ${\sim 10^4\, \makebox{V/sm}}$. 
Since the maximum strength is achieved near the sphere surface (i.e., at $r_q$), the value ${E_q=q/r_q^2}$ should be smaller than ${10^4\, \makebox{V/sm}}$. 

Let it be 
\begin{equation}
E_q = 10^3\, \makebox{V/sm} = 3.(3)\makebox{ <<CGSE>> units}
\label{E_q}\end{equation}
It follows from (\ref{r_q2}) that
\begin{equation}
E_q := \frac{q}{r_q^2} = \frac{\beta^{-2}\gamma^{-2}}{q}
\label{E_q2}\end{equation}
Based on (\ref{E_q2}), we can express the electrical charge $q$ as
\begin{equation}
q = \frac{\beta^{-2}\gamma^{-2}}{E_q}
\label{q1}\end{equation}
We assume that ${\beta^2 = 0.5}$ and ${\gamma = 0.5}$; then ${\kappa = \sqrt{8}/3}$. 
Having rewritten expression (\ref{q1}) in CGSE units, we will have (to the end of this section) the following expression for the charge in these units:
\begin{equation}
q = \frac{\beta^{-2}\gamma^{-2} c^4}{G E_q} \approx 2.915 \cdot 10^{49}\makebox{<<CGSE>> units} \approx 6.073 \cdot 10^{58}\makebox{<<e>>} 
\label{q2}\end{equation}
Correspondingly, for the total BH mass $M$ we have
\begin{equation}
M = \frac{q}{\kappa\sqrt{G}} \approx 1.197 \cdot 10^{53}\makebox{g} \approx 6.02 \cdot 10^{20}\, M_\odot 
\label{M1}\end{equation}
Based on (\ref{r_q2}) and (\ref{q2}), we find the charged sphere radius ${r_q}$ to be
\begin{equation}
r_q = \frac{q \beta\gamma \sqrt{G}}{c^2} = \frac{c^2}{\beta\gamma\sqrt{G}E_q} 
\approx \, 2.957\cdot 10^{24}\, \makebox{sm} \approx 958\, 000\makebox{pc} \sim 1\makebox{Mpc} 
\label{r_q3}\end{equation}
Let us estimate the density of the excess (or deficit) elementary charges comprising the charge $q$ of a sphere of radius $r_q$. 
According to (\ref{q2}) and (\ref{r_q3}), this surface charge density will be as low as 
${\sigma_q\sim 5.46\cdot 10^{8}\makebox{e/sm}^2\approx 8.74\cdot 10^{-7}\makebox{C/m}^2}$.

On the assumption that the charged-sphere matter has a density on the order of ${1\makebox{g/sm}^3}$, at ${q^2 = Gm_q^2}$, 
the thickness $d_q$ of this charged spherical shell, according to (\ref{q2}) and (\ref{r_q3}), turns out to be about ${70}$ m. 
As compared with the radius $r_q$ from (\ref{r_q3}), this is an ultrathin film; 
however, in comparison with the thickness of the layer in which excess (or deficit) electrons with a charge $q$ are concentrated 
(i.e., as compared with atomic sizes), this is an immense value.

One can also calculate the surface tension $N$ formed by the electrical force of repulsion for a surface charge $q$ on a sphere of radius $r_q$:
\begin{equation}
N = \frac{q^2}{4\pi r_q^3} = \frac{\beta^{-2}\gamma^{-2} c^4}{4\pi G r_q}
\approx 2.616\cdot 10^{24}\makebox{din/sm}
\label{N_1}\end{equation}
With the gravitational force of attraction neglected, we have to divide the tension N by the specific strength of the film material in order to calculate the minimum
thickness ${d_{min}}$ for a film retaining charge by its surface tension. 
Let this material be steel with a specific strength ${P_{s} = 10^{10}\makebox{d/sm}^2}$. 
Then we find from (\ref{N_1}) that
\begin{equation}
d_{min} = \frac{N}{P_{s}} = \frac{q^2}{4\pi P_{s} r_q^3} \approx 2.616\cdot 10^{14} \makebox{sm}
\label{d_min}\end{equation}
However, the mass of this steel will then be about 13 orders of magnitude larger than the total mass $M$, which is impossible. 
Thus, equilibrium cannot be reached by compensating the electrical force of repulsion with the elastic force of material. 

Note that the surface tension force is almost tangential to the small element of the charged-sphere surface. 
Therefore, to compensate for the radial electrical force, the surface tension force should greatly exceed the electrical force; the larger the radius ${r_q}$, 
the larger this excess must be. 

A calculation of the pressure $P$ exerted by the electrical force of repulsion on the charged-sphere surface yields
\begin{equation}
P = \frac{q^2}{4\pi r_q^4} \approx 0.88\, \makebox{d/sm}^2
\label{P_1}\end{equation}
At such large radii this is a very low pressure; practically all materials can withstand it. 
Such a large difference between the surface tension $N$ and pressure $P$ is due to the fact that ${N\propto 1/r_q^{3}}$, while ${P\propto 1/r_q^{4}}$; hence, ${P/N\propto 1/r_q}$. 
Therefore, if there is another radial force acting in the opposite direction (gravity force), this force can balance theoretically the electrical force (see below).

\section{POSSIBILITY OF EQUILIBRIUM ON A CHARGED SPHERE AT THE INITIAL MOMENT}
\label{s_equiv}

Let us consider the possibility of establishing equilibrium for a charged sphere of mass $m_q$, radius $r_q$, and charge $q$. 
This equilibrium will be maintained due to the compensation of the electrostatic repulsion of charges on the sphere 
by the sphere self-gravity. 
This equilibrium is possible only beyond the gravitational radius of the sphere. 
As will be seen from further calculations, the BH horizon may not form at all at these masses and charges.

It is known from the general-physics course that the electric field potential $A_t$ must be constant inside a
charged sphere and decrease inversely proportionally to radius beyond the sphere. 
Therefore, the electric field should be zero within the sphere and equal to ${q/r^2}$ beyond it. 
However, we will be interested in the electric field on the sphere itself. 
Obviously, it is equal to the arithmetic mean of the field limits within the sphere and beyond it when the radii tend to $r_q$; 
i.e., the electric field on the sphere is ${q/(2r_q^2)}$.

The same holds true for the gravitational charge (mass); 
however, we will assume that the mass is uniformly distributed over a layer of finite thickness $d$  
(the physical thickness of the sphere film). 
It will be shown below that the thickness $d$ should be much smaller than the sphere radius $r_q$; 
nevertheless, $d$ is a macroscopic size. 
At the same time, the charge $q$ is concentrated only on the sphere surface 
(in a layer whose thickness has atomic sizes). 
We assume that the charge is determined by the deficit of electrons on the sphere;
i.e., the positive charge of the nuclei of sphere lattice is not completely compensated by the elementary charge. 
The sizes of electrons and nuclei are much smaller than the sphere thickness $d$. 
Therefore, the gravitational field behaves identically at different sides of the sphere external surface, 
and the electric field under the external sphere surface becomes zero, in correspondence with the aforesaid.

We consider a charge element $e$ on the sphere and the corresponding mass element $\mu$, which obey the relations
\begin{equation}
e := \xi\cdot q \, , \quad \mu := \xi\cdot m_q \, ,\quad \xi = const << 1 \, .
\label{e_mu1}\end{equation}
Let us write now the equations of motion of a charged particle in gravitational and electromagnetic fields (see ~\cite{Landau1}, \S 90):
\begin{equation}
\mu\frac{du^i}{ds} = e F^{ij} u_j - \mu \Gamma^i_{km} u^k u^m \, ,\quad 
\Gamma^i_{km} = \frac{1}{2}g^{in}\left( \frac{\partial g_{nk}}{\partial x^m} + \frac{\partial g_{nm}}{\partial x^k} - \frac{\partial g_{km}}{\partial x^n} \right) \, .
\label{e_mu2}\end{equation}
Here, ${F_{ik}}$ is the electromagnetic field tensor, ${u^k}$ is
the 4-vector of velocity of a particle with a mass $\mu$ and a charge $e$, and $g_{in}$ are the components of the metric tensor.

In this section, we choose a static frame of reference and the Reissner–Nordström metric for it:
\begin{equation}
ds^2 = f dt^2 - f^{-1}\, dr^2\, -\, r^2\, (d\theta^2 + \sin^2\theta\, d\varphi^2) \, ,\quad 
f(r) := \left(1 - \frac{2m}{r} + \frac{q^2}{r^2} \right) \, .
\label{ds-0}\end{equation}
For this metric we have 
\begin{equation}
u^k = \delta^k_t u^t \, ,\quad u^t = f^{-1/2} \, ,\quad F_{ik} := \partial_i A_k - \partial_k A_i \, ,\quad A_i = \delta_i^t A_t\, .
\label{ds-0-1}\end{equation}
Here, ${A_i}$ is the 4-vector potential of electromagnetic field; 
${A_t = q/r}$ and ${A_t = q/r_q}$ at ${r\geq r_q}$ and ${r\leq r_q}$, respectively.

All quantities depend on only the radius, and Eq. (\ref{e_mu2}) can be rewritten in the form
\begin{equation}
\mu\frac{du^r}{ds} = -e \,\sqrt{f}\cdot F_{rt} - \frac{\mu}{2}\cdot\frac{df}{dr}
\label{e_mu3}\end{equation}
In correspondence with the aforesaid, we have the following expression for the single tensor component ${F_{ik}}$ on the sphere: 
${\left. F_{rt}\right|_{r_q} = -0.5q/r_q^2}$. 

In addition, a factor of ${0.5}$ should arise in the term with $q^2$ for the ${\partial_r f}$ in formula 
(\ref{e_mu3}), which can be rewritten for the equilibrium point in the form 
\begin{equation}
\left.\mu\frac{du^r}{ds}\right|_{r=r_q} = 0.5\cdot\frac{eq\sqrt{f(r_q)}}{r_q^2} 
- \frac{\mu}{2}\, \left(\frac{2m_q}{r_q^2} - 0.5\cdot\frac{2q^2}{r_q^3} \right) = 0
\label{e_mu3-2}\end{equation}
It follows from (\ref{e_mu3-2}) that this equilibrium point is stable, because the positive force of repulsion 
decreases as ${\propto 1/r_q^{3}}$, whereas the negative force of attraction decreases as${\propto 1/r_q^{2}}$.

\begin{figure*}
\centering
\includegraphics[width=0.5\textwidth]{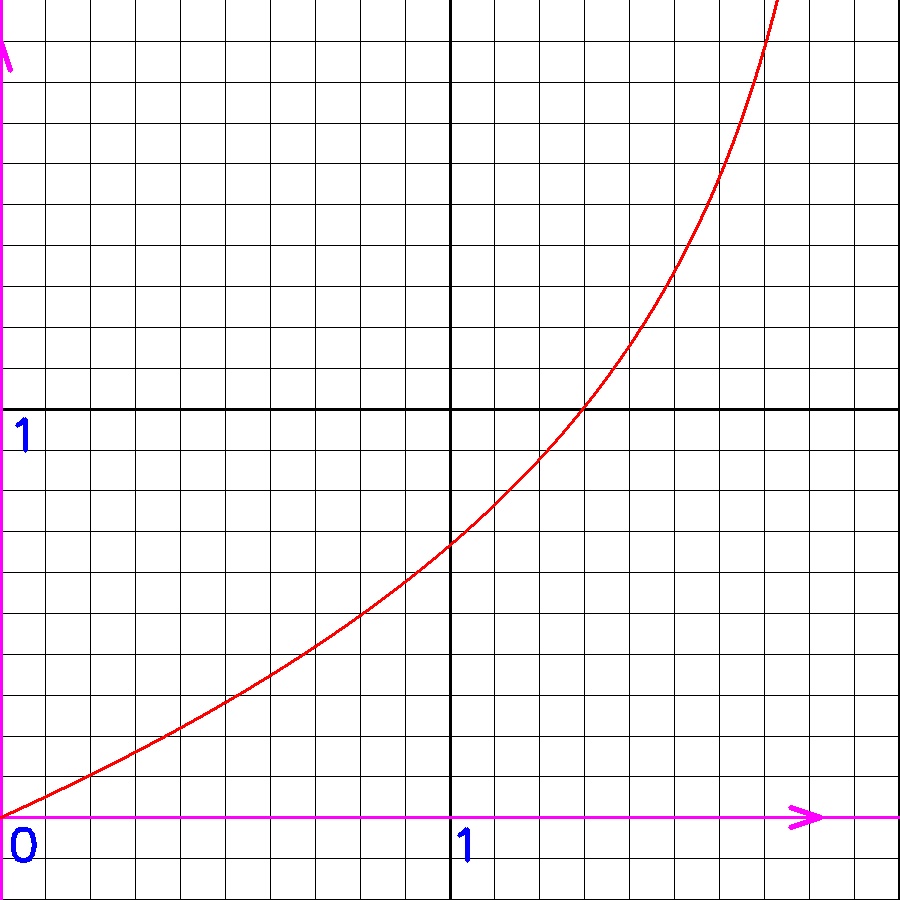}
\caption{{
Dependence of the charged-sphere radius $r_q$ on ${q^2}$ in units of mass $m_q$ (see (\ref{e_mu5})).
}}
\label{R1}\end{figure*}

Using expressions (\ref{e_mu1}), we replace the quantities ${(e,\,\mu)}$ with ${(q,\,m_q)}$ in Eq. (\ref{e_mu3-2}) and arrive at
\begin{eqnarray}
(4m_q^4 - q^4)r_q^2 -2q^2m_q(2m_q^2 - q^2)r_q + q^4(m_q^2 - q^2) = 0 \label{eq_rq} \\ 
r_q(q, m_q) = \frac{m_q q^2(2m_q^2 - q^2) + q^4\sqrt{2m_q^2 - q^2}}{4m_q^4 - q^4} \label{e_mu5}
\end{eqnarray}
As follows from (\ref{e_mu5}), the function ${r_q(q)}$ is increasing (see Fig.~\ref{R1}). 
According to (\ref{r_q2}), we are interested in only ${r_q<q}$; hence, the limitation on the charge $q$ takes the form
\begin{equation}
m_q^2 < q^2 < 2m_q^2
\label{e_mu6}\end{equation}
Therefore, when determining the root of $r_q$ in Eq. (\ref{eq_rq}), we chose the <<$+$>> sign. 
It follows from formula (\ref{e_mu6}) that a sphere with a mass  $m_q$ and charge $q$ cannot form a BH; 
otherwise, expressions (\ref{r_pm}) for horizons would be complex values.

\section{MODEL EQUATIONS}
\label{s_eq}

The frame of reference comoving synchronously with matte\footnote{
The frame of reference synchronously comoving with matter is applicable to dust matter, because pressure is absent in dust 
(the metric time component is 1).}, 
described by coordinates $\tau$ and $R$,
\begin{equation}
ds^2 = d\tau^2 - e^{\lambda (\tau ,R)}\, dR^2\, -\, r^2(\tau ,R)\, (d\theta^2 + \sin^2\theta\, d\varphi^2) ,
\label{ds-1}\end{equation}
turns out to be more convenient for further analysis. 
Let the accreting matter be a gravitating dust with an energy density ${\varepsilon}$, whose distribution depends on
only the radius $r$. 
We will also take into account the influence of the cosmological $\Lambda$-term on the matter 
dynamics in our model equations. 
Then the total energy–momentum tensor beyond a sphere of radius $r_q$ can be presented in the form
\begin{equation}
T^n_m=\left(
\begin{tabular}{c c c c}
$+\frac{q^2}{8\pi r^4}$ & 0 & 0 & 0 \\
0 & $+\frac{q^2}{8\pi r^4}$ & 0 & 0 \\
0 & 0 & $-\frac{q^2}{8\pi r^4}$ & 0 \\
0 & 0 & 0 & $-\frac{q^2}{8\pi r^4}$ \\
\end{tabular}
\right)+\left(
\begin{tabular}{c c c c}
$\varepsilon$ & 0 & 0 & 0 \\
0 & 0 & 0 & 0 \\
0 & 0 & 0 & 0 \\
0 & 0 & 0 & 0 \\
\end{tabular}
\right) 
+ \delta^n_m \Lambda
\label{2-2}\end{equation} 
The first term on the right-hand side of (\ref{2-2}) corresponds to energy–momentum tensor of the electromagnetic field 
of a point charge $q$, 
the second term corresponds to the energy–momentum tensor of dust, and the term 
${\delta^n_m \Lambda}$ is the cosmological ${\Lambda}$-term. 
Inside a sphere of radius $r_q$ the energy–momentum tensor of electromagnetic field should be zero, 
because electric field is absent in a charged sphere. 
Concerning the energy–momentum tensor of electromagnetic field on the sphere of radius $r_q$, 
in correspondence with the considerations reported in Section~\ref{s_equiv}, 
all tensor components on this sphere must be divided by 2, 
because the electric field on the sphere is an average of the fields inside the sphere and beyond it.

Due to the hydrodynamic independence of dust layers relative to each other, one can integrate the 
equations of motion for dust similarly to the solution of Tolman’s problem (see~\cite{Tolman, Oppenheimer}). 
In essence, this is the same Tolman’s problem in a centrally symmetric electric field for neutral dust. 

The Einstein equations corresponding to metric (\ref{ds-1}) can be written as\footnote{
Derivation of Eqs. (\ref{En1})-(\ref{En4}) can be found, e.g., in \cite{Landau1} (\S 100, problem 5).}: 
\begin{eqnarray}
8\pi T_\tau^\tau=8\pi\varepsilon +q^2/r^4 + 8\pi\Lambda = 
\frac{1}{r^2}\left[1+r r_{,_\tau}\lambda_{,_\tau}+r^2_{,_\tau}-
e^{-\lambda}\left( 2r r_{,_{RR}} + r^2_{,_R}- r r_{,_R}\lambda_{,_R} \right) 
\right]
\label{En1}\\ 
8\pi T_R^R = q^2/r^4 + 8\pi\Lambda = 
\frac{1}{r^2}\left( 1+ 2r r_{,_{\tau\tau}} + r^2_{,_\tau}   
- e^{-\lambda} r^2_{,_R}\right)
\label{En2}\\ 
8\pi T^R_\tau=0=\left( 2r_{,_{R\tau}}-r_{,_R}\lambda_{,_\tau}\right) 
e^{-\lambda}/r 
\label{En3}\\ 
8\pi T^\theta_\theta = 8\pi T^\varphi_\varphi = -q^2/r^4 + 8\pi\Lambda =
\frac{r_{,_{\tau\tau}}}{r}+\frac{\lambda_{,_{\tau\tau}}}{2}+
\frac{\lambda_{,_\tau}^2}{4}+\frac{r_{,_\tau}\lambda_{,_\tau}}{2r}
-\left( r_{,_{RR}} - \frac{1}{2}r_{,_R}\lambda_{,_R}\right) e^{-\lambda}/r  
\label{En4}\end{eqnarray}
Integrating Eq. (\ref{En3}) over time, we obtain
\begin{equation} 
e^{-\lambda}r^2_{,_R}=F_1(R) 
\label{3-1}\end{equation}
The function ${F_1(R)}$, as will be seen below, determines the initial conditions for the dust velocity distribution. 

Substituting (\ref{3-1}) into Eq. (\ref{En2}), we find that
\begin{equation}
\frac{q^2}{r^2} + 8\pi\Lambda r^2 = \left( r r^2_{,_\tau} \right)_{,\tau}/r_{,_\tau} -F_1+1 
\label{3-2}\end{equation}
Multiplying this equation by $r_{,\tau}$ and integrating the result over time, we arrive at
\begin{equation}
\frac{q^2}{r} - \frac{8\pi}{3}\Lambda r^3 + r (1-F_1) + r r^2_{,_\tau} = F_2(R) 
\label{3-3}\end{equation}
The function ${F_2(R)}$, as will be seen below, determines the initial conditions for the dust density distribution.

\section{INITIAL CONDITIONS}
\label{s_init}

Furthermore the subscript <<i>> will be assigned to all quantities at ${\tau =0}$. 

We choose the scale of coordinate $R$ in (\ref{ds-1}) at the initial instant such that
\begin{equation}
r_i := R 
\label{ini2}\end{equation}
This can be done if the function r changes monotonically along the coordinate $R$ at the initial instant. 

We assume also that the entire matter is at rest at the initial instant and that all derivatives with respect to $\tau$ are zero:
\begin{equation}
\left. r_{,_\tau}\right|_i = 0
\label{ini3}\end{equation}
Then, proceeding from (\ref{3-3}), we obtain
\begin{equation}
F_2(R) = \frac{q^2}{R} - \frac{8\pi}{3}\Lambda R^3 + R (1-F_1) 
\label{F_2}\end{equation}
or
\begin{equation} 
F_1(R) = 1 - \frac{F_2}{R} + \frac{q^2}{R^2} - \frac{8\pi}{3}\Lambda R^2
\label{F_1}\end{equation}
Multiplying Eq. (\ref{En1}) by ${r^2 r_{,_R}}$ and expressing ${\lambda_{,_\tau}}$ from Eq. (\ref{En3}), we arrive at
\begin{equation}
8\pi\varepsilon r^2 r_{,_R} = \left[q^2/r - \frac{8\pi}{3}\Lambda r^3 + r(1-F_1) + r r_{,_\tau}^2\right]_{,_R} 
\label{En1-2}\end{equation} 
A comparison of this expression with (\ref{3-3}) shows that
\begin{equation}
8\pi\varepsilon r^2 r_{,_R} = \frac{dF_2}{dR}
\label{En1-2-2}\end{equation} 
Based on Eq. (\ref{En1-2-2}), we can conclude that the quantity ${8\pi\varepsilon r^2 r_{,_R}}$ is independent of time:
\begin{equation}
8\pi\varepsilon r^2 r_{,_R} = 8\pi\varepsilon_i r_i^2 
\label{bianki1}\end{equation}
Let us introduce the mass function ${m(R)}$: 
\begin{equation} 
m(R) := \int\limits_0^R 4\pi\varepsilon_i R^2 \, dR = \int\limits_0^{r_i} 4\pi\varepsilon r^2 \, dr
\label{3-6}\end{equation}
Within our model, there is neither matter nor electric field at radii smaller than $r_q$  
(except for the $\Lambda$-term), and there is a charged thin film with a mass of $m_q$ at the radius $r_q$. 
Therefore\footnote{The superscripts <<${-}$>> and <<${+}$>> for the radius $r_q$ indicate the lower and upper limits, i.e., 
respectively, under and above the charged film with a mass $m_q$.}, 
${m(r_q^-) \equiv 0}$ и ${m(r_q^+) \equiv m_q}$ and expression (\ref{3-6}) can be integrated not from zero 
but from ${r_q^-}$ or from ${r_q^+}$: 
\begin{eqnarray}
m(R>r_q^-) = \int\limits_{r_q^-}^R 4\pi\varepsilon_i R^2 \, dR \label{m_Rm} \\
m(R>r_q^+) = m_q + \int\limits_{r_q^+}^R 4\pi\varepsilon_i R^2 \, dR \label{m_Rp}
\end{eqnarray}
Integrating expression (\ref{En1-2-2}) in the limits from ${r_q^-}$ to $R$ and from ${r_q^+}$ to $R$, 
we find from (\ref{m_Rm}-\ref{m_Rp}) that 
\begin{eqnarray}
2m(R>r_q^-) = F_2(R>r_q^-) - F_2(r_q^-) \label{5-mass3} \\
2m(R>r_q^+) - 2m_q = F_2(R>r_q^+) - F_2(r_q^+) \label{5-mass2} 
\end{eqnarray}
To obtain the $F_1$ and $F_2$ values at the point $r_q$, we will take into account that the charged sphere is in 
equilibrium; therefore, the following equalities must be fulfilled: ${r_{,_\tau}=0}$ и ${r_{,_{\tau\tau}}=0}$. 
In addition, the electric field is absent under the charged sphere (at $r_q^-$) 
(i.e., the terms with $q^2$ must be excluded in that domain), whereas directly above the charged sphere the electric 
field is ${q/r_q^2}$; therefore, the terms with $q^2$ on the charged sphere must be multiplied by ${\frac{1}{2}}$. 
Then, proceeding from Eqs. (\ref{En2}), (\ref{3-1}) and (\ref{F_2}), we find the ${F_1}$ and ${F_2}$ values on the charged sphere:
\begin{eqnarray}
F_1(r_q) = 1 - \frac{q^2}{2r_q^2} - 8\pi\Lambda r_q^2 \label{F_1q} \\
F_2(r_q) = \frac{q^2}{r_q} + \frac{16\pi\Lambda}{3} r_q^3 \label{F_2q}
\end{eqnarray}
As was noted in Section~\ref{s_desc}, the matter mass ${m_q}$ has accumulated completely at the charge radius; 
therefore, expressions for ${r_q^+}$ must be used in the calculations. 
Then, proceeding from (\ref{m_Rp}), (\ref{5-mass2}) and (\ref{F_2q}), we obtain the following expression for 
${F_2(R>r_q^+)}$: 
\begin{eqnarray}
F_2(R>r_q^+) = \frac{q^2}{r_q} + \int\limits_{r_q^+}^R 8\pi\varepsilon_i R^2 \, dR + \frac{16\pi\Lambda}{3} r_q^3 
= \frac{q^2}{r_q} + 2m_{dust}(R) + \frac{16\pi\Lambda}{3} r_q^3 
\label{F_2_2}\end{eqnarray}
Let us introduce also the mass ${m^\Lambda(R)}$, related to the $\Lambda$-term:
\begin{equation} 
m^\Lambda(R) := \int\limits_0^{R} 4\pi\Lambda R^2 \, dR = \frac{4\pi}{3}\Lambda R^3 
\label{delta_m_q}\end{equation}
It follows from (\ref{3-6}) that the consideration of the cosmological ${\Lambda}$-term can be reduced to subtraction of 
${4\pi\Lambda R^3/3}$ from the mass${m(R)}$. 
According to the cosmological observations, ${\Lambda\sim 10^{-29}\makebox{g/sm}^3}$; therefore, on 
the scales on the order of $r_q$, which are determined by expression (\ref{r_q3}), 
the mass related to the $\Lambda$-term is about 8 orders of magnitude smaller than the total mass $M$ (see~(\ref{M1})). 
This means that the cosmological $\Lambda$-term can be disregarded at ${R\sim r_q}$.

\section{INTEGRATION OF THE EQUATIONS}
\label{s_int}

The horizon radii ${r_h^\pm}$ in comoving coordinates are determined by the equality of the modulus of invariant 
velocity $V$ to unity:
\begin{equation} 
\left. V^2\right|_{r_h^\pm} = 1 \, ,\quad 
V^2(\tau, R) := \frac{r_{,\tau}^2\, e^\lambda}{r_{,R}^2} \, .
\label{V2}\end{equation}
Here, the function ${V^2(\tau, R)}$ is the squared invariant velocity (see~\cite{Shatsk1, Shatskiy2010}). 
Using expression (\ref{3-1}), one can rewrite Eq. (\ref{V2}) in the form 
\begin{equation} 
\left.r_{,\tau}^2\right|_{r_h^\pm} = F_1(R)
\label{V2_0}\end{equation}
Having expressed (using (\ref{F_1})) the function $F_1$ in Eq. (\ref{3-3}) in terms of $F_2$, 
we rewrite (\ref{3-3}) in the form convenient for integrating and obtaining the function ${r(\tau, R)}$:
\begin{equation} 
r_{,_\tau} = \pm\sqrt{\left(\frac{F_2}{r} - \frac{q^2}{r^2} + \frac{8\pi\Lambda r^2}{3}\right) - 
\left(\frac{F_2}{R} - \frac{q^2}{R^2} + \frac{8\pi\Lambda R^2}{3}\right)}
\label{r_get}\end{equation}
This expression demonstrates, in particular, that the particle dynamics depends on only the distribution 
of matter under the radius determining the particle location and is independent of matter above this radius.

Proceeding from (\ref{r_get}) and taking into account (\ref{F_1}) and (\ref{V2_0}), 
we obtain an equation for calculating the horizon radii ${r_h^\pm}$:
\begin{equation} 
1 - \frac{F_2}{r_h^\pm} + \frac{q^2}{{r_h^\pm}^2} - \frac{8\pi}{3}\Lambda {r_h^\pm}^2  = 0 
\label{V2_1}\end{equation}
This expression, with the $\Lambda$-term disregarded, is practically equivalent to the equality ${f(r_h^\pm)=0}$ 
from definition (\ref{ds-0}); 
i.e., the meaning of function $F_2$ is the doubled mass $m$ in the matter-comoving coordinates.
Therefore, we neglect below the $\Lambda$-term and derive an analogue of expression (\ref{V2_1}) 
from (\ref{r_pm}) for the horizon radii in matter-comoving coordinates:
\begin{equation} 
r_h^\pm = \frac{1}{2}F_2 \pm \sqrt{\frac{1}{4}F_2^2 - q^2} 
\label{hor_pm}\end{equation}
Hence, in correspondence with (\ref{r_qM}), we have
\begin{equation}
q = \frac{\kappa F_2}{2} \, ,\quad \kappa := \frac{2\beta}{1+\beta^2} \, .
\label{r_qF2}\end{equation}  
Based on Eq. (\ref{r_get}), one can obtain two stopping points for the dust shell; 
these are turning points, at which ${r_{,_\tau} = 0}$. 
Expression (\ref{r_get}) yields directly the trivial stopping point: 
the initial point ${r_i := R}$, at which the dust shell begins to move. 
Concerning the desired second stopping point, we find from (\ref{r_get}) that
\begin{eqnarray}
r_{t} = \frac{q^2}{F_2 - q^2/R} = \frac{\kappa q}{2 - \kappa q/R}
\label{r_t1}\end{eqnarray}
Hence, one can see that ${r_{t} \ge 0.5\kappa q}$; the equality is implemented only at ${q/R\to 0}$.

Let us first consider the motion of probe particles under an external dust shell. 
Probe particles cannot form a horizon; 
according to (\ref{F_2_2}) and (\ref{r_t1}), we have the following relations for them:
\begin{eqnarray}
F_2^0 = \frac{q^2}{r_q} \, ,\quad r_{t}^0 = \frac{r_q R}{R - r_q} \, .
\label{r_t0}\end{eqnarray}
Therefore, in the case ${r_q<R<2r_q}$ for probe particles, 
the stopping radius $r_{t}^0$ even exceeds the initial radius $r_i$, 
which indicates the outside direction of their motion. 
Correspondingly, in the range ${2r_q<R<\infty}$ the probe particles will begin to move inside. 
For the stopping point of a probe particle with ${r_{t}^0 \to r_q}$, 
we find from (\ref{r_t0}) that ${R\to\infty}$. 
Thus, the spherical layers of probe particles will constantly cross each other during their motion.

\begin{figure}
\centering
\begin{minipage}[h]{0.49\linewidth}
\center{\includegraphics[width=0.90\linewidth]{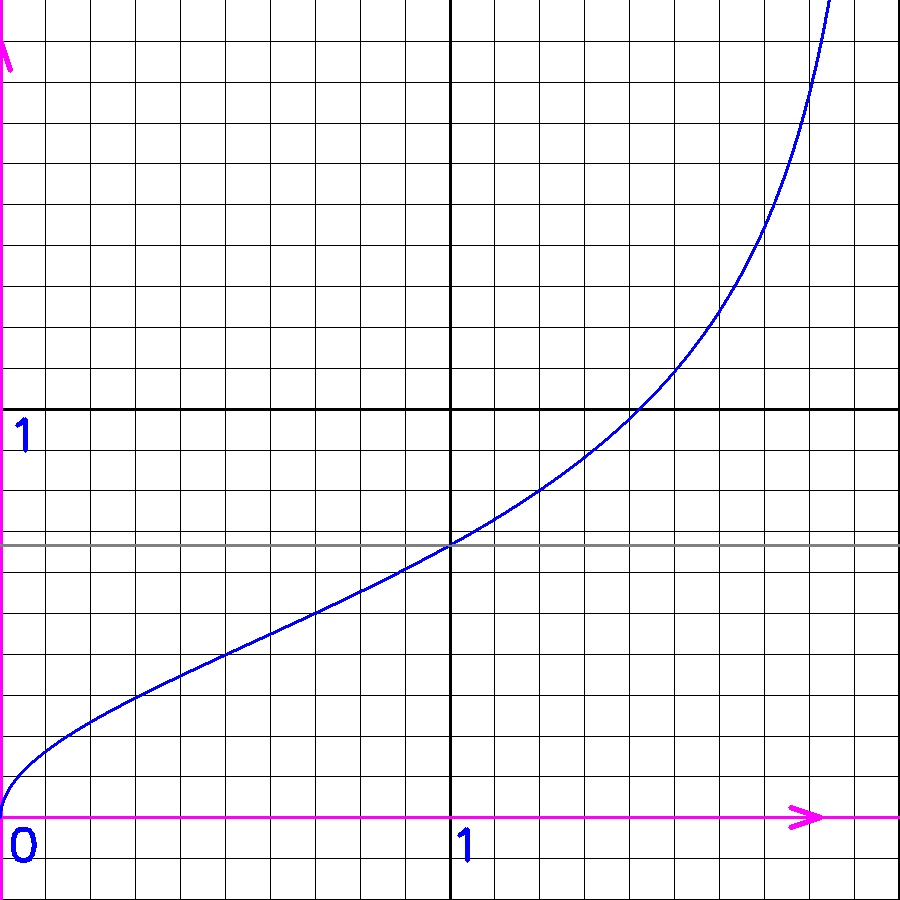}}
\end{minipage}
\begin{minipage}[h]{0.49\linewidth}
\center{\includegraphics[width=0.90\linewidth]{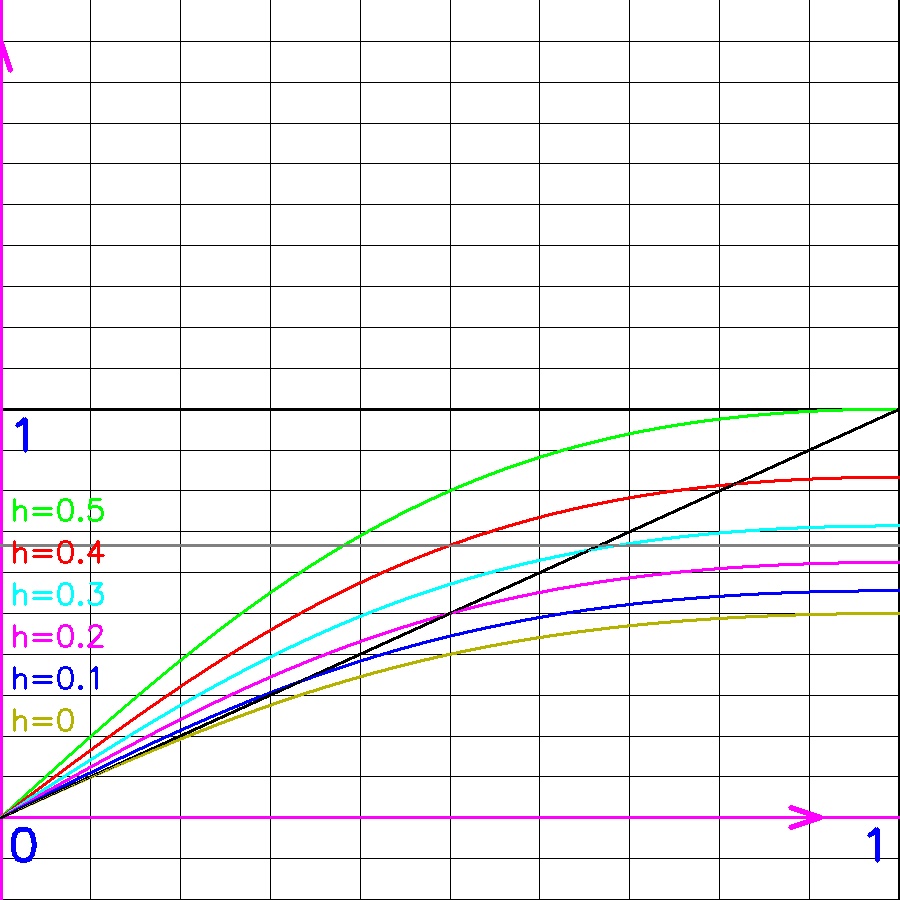}}
\end{minipage}
\caption{{
(Left) Dependence of ${r_q/q}$ on ${q^2/m_q^2}$. 
\newline
(Right) Dependences of ${r_t/q}$ on $\beta$ at different $h$ for inequality (\ref{r_b1_5}). 
\newline
.\hrulefill
}}
\label{R2}
\end{figure}

Let us now consider the non-probe particles of a dust sphere of mass ${m_{dust}}$, 
whose radius is $R_1$ at the initial instant. 
In accordance with (\ref{r_q2}), the condition for location of turning point radius 
$r_t$ for dust-sphere particles 
between the charge-sphere radius $r_q$ and Cauchy horizon radius $r_h^-$ takes the form
\begin{eqnarray}
r_q = \beta\gamma q < r_{t} < r_h^{-} = \beta q \, ,\quad m_q < q \, .
\label{r_b1_3}\end{eqnarray}
Introducing the designation 
\begin{eqnarray}
h := \frac{\kappa q}{2R_1} 
\label{r_b1_4}\end{eqnarray}
and taking into account (\ref{e_mu5}) and (\ref{r_t1}), we rewrite (\ref{r_b1_3}) as
\begin{eqnarray}
\frac{r_q}{q} = \frac{m_q q(2m_q^2 - q^2) + q^3\sqrt{2m_q^2 - q^2}}{4m_q^4 - q^4}
 < \frac{r_t}{q} = \frac{\beta}{\left(1+\beta^2\right)\left(1 - h\right)} < \beta 
\, ,\quad \frac{q}{m_q} > 1 \, .
\label{r_b1_5}\end{eqnarray}
The range of parameters at which these inequalities are satisfied can be determined from the plots presented 
in Fig.~\ref{R2}.

Here, we should also consider a seeming contradiction related to possible intersection of dust layers. 
As was shown for probe particles, they do not reach the radius $r_q$ at finite $R$ values. 
The same holds true for the dust-shell inner layers, because the gravitational 
potential for these layers is still insufficient even to form BH horizons. 
Therefore, the dust-shell inner layers will begin to move from the center at some instant. 
At the same time, the dust-shell outer layers have already a sufficiently high gravitational potential to 
form BH horizons and reach the turning point ${r_{t}}$. 
Thus, the inner layers will cross the outer layers and their roles will interchange at some instant. 
Theoretically, our model becomes incorrect after the intersection 
of non-probe dust layers; however, we deal in practice with a fairly thin dust shell, whose thickness 
was chosen by us incommensurately smaller than the radius $r_q$ (see~(\ref{r_q3})). 
Therefore, both under and above this thin dust shell, we can neglect the inner intersection 
of its layers, whereas inside the dust shell there will be constant mixing of its inner and outer layers. 
On the whole, this pattern should not affect the model dynamics. 
In addition, we can slightly modify our model and assume that bulk dust shells consist of only one layer; 
the size of each <<dust particle>> in this layer is on the order of the entire dust shell thickness ${d_{dust}}$ 
(${d_{dust}<<r_q}$). 
In this case, there will be no intersection of dust layers, and each bulk <<dust particle>> will move 
in a self-consistent gravitational field. 
According to the estimates reported in Section~\ref{s_desc}, ${m_{dust}\sim m_q\sim M}$ in our model; 
therefore, at a density of <<dust particle>> material of about 
${\sim 1\makebox{g/sm}^3}$, one can obtain a limitation 
on the <<dust particle>> size: ddust ${d_{dust}\lesssim d_q\approx 70}$. 
Therefore, the <<dust particles>> in the dust layer under consideration can be small asteroids or boulders.

\section{CHOICE OF APPROPRIATE PARAMETERS}
\label{s_podbor}

Taking into account (\ref{e_mu6}), we introduce the following designation:
\begin{eqnarray}
\alpha := \frac{q}{m_q} > 1
\label{r_q_eq4}\end{eqnarray}
To implement the electrostatic and gravitational equilibrium of charge $q$ on a sphere with a mass $m_q$ and
radius $r_q$, one must choose the parameters $\alpha$, $\beta$ and $h$ such as to satisfy inequalities 
(\ref{r_b1_5}). 
This can be done using, e.g., the parameters 
\begin{eqnarray}
\alpha^2 = 1.1 \, ,\quad \beta = 0.9 \, ,\quad h = 0.4 \nonumber \\
\Rightarrow\, \frac{r_q}{q}\approx 0.731<\frac{r_t}{q}\approx 0.829\, ,\quad \kappa \approx 0.997 \, ,\quad \frac{R_1}{r_q} \approx 1.706 \, .
\label{par1}\end{eqnarray}
or  
\begin{eqnarray}
\alpha^2 = 1.01 \, ,\quad \beta = 0.8 \, ,\quad h = 0.3 \nonumber \\
\Rightarrow\, \frac{r_q}{q}\approx 0.673<\frac{r_t}{q}\approx 0.697\, ,\quad \kappa \approx 0.988 \, ,\quad \frac{R_1}{r_q} \approx 2.447 \, .
\label{par2}\end{eqnarray}
Thus, the problem of calculating the equilibrium state of charge $q$ on a sphere can be considered as solved.

\section{ACCRETION OF MATTER ON ALREADY EXISTING REISSNER–NORDSTRÖM BH}
\label{s_bh}

We considered above the accretion of a dust shell on a charged sphere in equilibrium, in the absence of BH. 
Let us now consider the accretion of a dust shell on a Reissner–Nordström BH. 

In accordance with (\ref{r_pm}), any charged BH with a mass $M$ and charge $q$ obeys the inequality
\begin{equation}
r_h^- \leq q \leq M \leq r_h^+
\label{r_pm2}\end{equation}
Hence, if a BH already exists at the initial instant, its charge remains constant, and only its mass 
increases, the following inequality will always be valid for any two instants <<$a$>> and <<$b$>>: 
\begin{equation}
r_{ha}^{-} \leq r_{hb}^{+}
\label{r_pm2}\end{equation}
Here, the subscripts <<$a$>> and <<$b$>> correspond to the radii of horizons the BH will have at the corresponding 
(arbitrary) instants. In other words, the case where the inner Cauchy horizon at some instant <<$a$>> 
exceeds the BH outer horizon at some other instant <<$b$>> is impossible. 
Hence, independent of the specific features of neutral matter accretion on the BH, 
the latter will have only one turning point. 
Therefore, the situation where another turning point would arise above the initial BH 
horizon (as a result of the formation of another outer $T$-region) is impossible.

\section{DISCUSSION}
\label{s_discuss}

Thus, the formation of another turning point for a charged BH is impossible after the formation instant 
of this BH. 
Therefore, a necessary initial condition for the turning point formation is the absence of BH at the 
initial instant. 
Hence, the occurrence of a new BH and a new universe 
(along with the occurrence of a white hole) is possible only at the instant when horizons of a 
new Reissner–Nordström BH (or, possibly, Kerr BH) arise.

Therefore, the above-considered mechanism of turning-point formation for a Reissner–Nordström 
BH is the only possible one for this BH. 
In other words, within our model, there is a charged sphere at the initial instant; 
the charge and entire matter of this sphere are concentrated in a thin charged spherical film. 
Beyond this sphere, there is no matter, up to some radius ${r_i=R_1}$. 
At this radius, $R_1$, a spherical dust shell with a mass ${m_{dust}}$ is at rest; 
the thickness ${m_{dust}}$ of this sphere is neglected (in comparison with the radius $r_q$). 
Horizons are absent at the initial instant (there is no BH). 
We are not interested in the matter beyond the radius ${r_i=R_1}$.

At subsequent instants the dust sphere begins to fall to the center (towards the charged sphere) under the 
action of gravity. 
At some instant the dust sphere reaches its gravitational radius ${r_h^{+}}$ (see~(\ref{hor_pm})) and finds 
itself under the horizon of newly formed Reissner–Nordström BH at the next instant. 

When the dust sphere reaches the turning-point radius, according to (\ref{V2}), 
its velocity $V$ decreases to zero, after which the dust sphere bounces to another universe. 
This is due to the fact that the dust sphere is already located in the inner R region 
(under both BH horizons). 
Therefore, the dust sphere cannot escape backwards, and the solution indicates that its radius 
should increase after the bounce. 
The only possibility is the escape of the dust sphere to another universe. 
This event occurs at the instant of the formation of a new white hole and new expanding universe. 
Apparently, the universe to which the dust sphere escapes arises jointly with the formation 
of both horizons of Reissner–Nordström BH. 

One of the most interesting conclusions that can be drawn is that the falling dust sphere in 
the model under consideration does not reach the charged sphere of radius $r_q$. 
The reason is that the turning point for the dust sphere with $r_{t}$ is located beyond the charged sphere. 
Therefore, a Reissner–Nordström black-andwhite hole with all its horizons and R-T-regions arises 
beyond the charged sphere. 
Apparently, nothing should change within the charged sphere of radius $r_q$: 
the electric field and singularity are absent even after the formation of a Reissner–Nordström BH outside 
the sphere. 
The metric coefficients in the model should remain everywhere such as to provide constant 
and continuous matching of two regions: within the charged sphere and beyond it. 

Thus, we proposed a model in which both the black-and-white hole and its turning point are formed 
in already existing Universe; 
i.e., the black-and-white hole in this model is not <<eternal>>. 
The analytical solution of a similar self-consistent problem of accretion 
for a real rotating Kerr BH would be practically impossible in view of its complexity. 
However, one can suggest that a new Kerr black-and-white hole is 
formed according to a similar mechanism during the rotating-star collapse. 
An important difference from the dynamic Reissner–Nordström solution is that the 
Kerr solution does not contain any limitations (see Section~\ref{s_desc}) on the electric field magnitude and,  correspondingly, on the sizes and mass of black-and-white hole. 
The specific features of the arising complex topology in a Kerr black-and-white hole were considered, 
e.g., in~\cite{Shatskiy2020, Carter1966, Carter1968}.

\hrulefill

ACKNOWLEDGMENTS 

I am grateful to the organizers and all members of Zelmanov Memorial Seminars in Gravitation and Cosmology 
at the Shternberg State Astronomical Institute of Moscow State University for the questions, 
discussion, and fruitful remarks.

\end{document}